\documentclass[aps,pra,showpacs,floatfix,twocolumn,superscriptaddress,bibnotes]{revtex4}
\usepackage{graphicx}
\usepackage{amsmath}
\usepackage{bm}
\usepackage{color}
\usepackage{dcolumn}
\usepackage{hyperref}

\begin{document}


\newcommand{\cmt}[1]{\textbf{[\![#1]\!]}}
\newcommand{\etal}{{\it et al.}}
\newcommand{\cm}{cm$^{-1}$}
\newcommand{\pprime}{{\prime\prime}}
\newcommand{\ls}[4]{\ensuremath{^{#1}\!{#2}_{#3}^{#4}}}
\newcommand{\jj}[4]{\ensuremath{\left({\textstyle #1,#2}\right)_{#3}^{#4}}}
\newcommand{\s}{\ls{1}{S}{0}{}}
\newcommand{\p}{\ls{3}{P}{0}{\circ}}
\newcommand{\yb}[1]{$^{#1}$Yb}
\newcommand{\clocks}{\ensuremath{6s^2\,\ls{1}{S}{0}{}}}
\newcommand{\clockp}{\ensuremath{6s6p\,\ls{3}{P}{0}{\circ}}}
\newcommand{\sr}[1]{$^{#1}$Sr}

\newcommand{\NIST}{
National Institute of Standards and Technology,
325 Broadway, Boulder, Colorado 80305, USA}

\newcommand{\CU}{
University of Colorado, Department of Physics,
Boulder, Colorado 80309, USA}

\newcommand{\NISTMD}{
National Institute of Standards and Technology,
100 Bureau Drive, Gaithersburg, Maryland 20899, USA}

\title{An atomic clock with $1\times 10^{-18}$ room-temperature blackbody Stark uncertainty}

\author{K. Beloy}
\email{kyle.beloy@nist.gov}
\affiliation{\NIST}
\author{N. Hinkley}
\affiliation{\NIST}
\affiliation{\CU}
\author{N. B. Phillips}
\affiliation{\NIST}
\author{J. A. Sherman}
\affiliation{\NIST}
\author{M. Schioppo}
\affiliation{\NIST}
\author{J. Lehman}
\affiliation{\NIST}
\author{A. Feldman}
\affiliation{\NIST}
\author{L. M. Hanssen}
\affiliation{\NISTMD}
\author{C. W. Oates}
\affiliation{\NIST}
\author{A. D. Ludlow}
\email{andrew.ludlow@nist.gov}
\affiliation{\NIST}

\date{\today}

\begin{abstract}
The Stark shift due to blackbody radiation (BBR) is the key factor limiting the performance of many atomic frequency standards, with the BBR environment inside the clock apparatus being difficult to characterize at a high level of precision.  Here we demonstrate an in-vacuum radiation shield that furnishes a uniform, well-characterized BBR environment for the atoms in an ytterbium optical lattice clock. Operated at room temperature, this shield enables specification of the BBR environment to a corresponding fractional clock uncertainty contribution of $5.5 \times 10^{-19}$.  Combined with uncertainty in the atomic response, the total uncertainty of the BBR Stark shift is now $1\times10^{-18}$.  Further operation of the shield at elevated temperatures enables a direct measure of the BBR shift temperature dependence and demonstrates consistency between our evaluated BBR environment and the expected atomic response.
\end{abstract}


\maketitle

The ability to control quantum systems facilitates their study and use in precision measurement experiments.  This is exemplified by the most advanced atomic clocks, which prepare trapped ultracold atoms in a single quantum state, resonantly drive these atoms with ultra-coherent laser or microwave fields, and detect the atomic state with high fidelity. Careful control of these quantum systems also requires minimizing or stabilizing perturbative influences that affect the internal atomic structure being probed.  The Stark shift due to blackbody radiation (BBR) constitutes one of the largest perturbations to the clock transition frequency of many atomic clocks, including cesium fountains, single-trapped ion systems, and optical lattice clocks. Consequently, the uncertainty stemming from this shift has played a dominant role in the total uncertainty of these standards (e.g.,~Refs.~\cite{Jefferts2014,Lemke09,Ludlow08,Huntemann2012,Madej2012,Falke2014,Bloom2014}).
The BBR shift can be expressed concisely as \cite{Porsev2006}
\begin{equation}
\Delta \nu_\mathrm{BBR} = - \frac{1}{2} \frac{\Delta \alpha(0)}{h} \langle E^2 \rangle_T \left[1 + \eta_\mathrm{clock}(T)\right],
\label{Eq:DeltanuBBR}
\end{equation}
where $\Delta \alpha(0)$ is the differential static polarizability between the two clock states, $h$ is Planck's constant, $\langle E^2 \rangle_T = \left[8.319430(15)\mathrm{~V/cm}\right]^2 (T/300\mathrm{~K})^4$ is the mean-squared electric field in a BBR environment of absolute temperature $T$~\cite{Angstmann2006}, and $\eta_\mathrm{clock}(T) \approx \eta_1 (T/300\mathrm{~K})^2 + \eta_2 (T/300\mathrm{~K})^4$ provides a small dynamic correction to account for frequency-dependence of the state polarizabilities across the BBR spectrum.
Evaluation of $\Delta \nu_\mathrm{BBR}$ requires (i) knowledge of the atomic response to BBR, as given by $\Delta \alpha(0)$ and $\eta_{1,2}$, as well as (ii) knowledge of the BBR environment, as given by the temperature $T$.

Recent efforts have improved knowledge of the atomic response to BBR in many atomic clocks.  For example, precise evaluation of $\Delta \alpha(0)$ and $\eta_{1,2}$ in optical lattice clocks based on Yb and Sr have reduced clock uncertainty from atomic response from the $10^{-16}$ fractional level to $\lesssim\!10^{-18}$ in the past few years \cite{Sherman2012,Middelmann2012,Beloy2012,Safronova2012,Safronova2013}.
In the Sr$^+$ clock, measurements exploiting time dilation of the trapped ion improved knowledge of $\Delta \alpha(0)$, resulting in a clock uncertainty due to atomic response below $10^{-18}$ \cite{Dube2014}.  In all these cases, uncertainty due to imprecise knowledge of the BBR environment had remained $\geq\!10^{-17}$.  Temperature inhomogeneities in the clock apparatus inevitably lead to deviations from an ideal BBR environment, requiring additional care in the interpretation of $T$ contained in $\langle E^2 \rangle_T$ and $\eta_\mathrm{clock}(T)$. Efforts to reduce and monitor temperature inhomogeneities in the vacuum chamber surrounding the atoms in a Sr lattice clock led to a BBR shift uncertainty of $4\times10^{-17}$ \cite{Falke2014}.  More recently, other strategies have sought to improve this further.  One approach directly samples the local radiation environment within a room-temperature apparatus, reporting a BBR shift uncertainty of $4\times10^{-18}$ for a Sr optical lattice clock \cite{Bloom2014}.  The other involves operation of the clock in a cryogenic environment where the BBR shift is suppressed.  A Cs fountain with a cryogenically-cooled microwave cavity and fountain chamber recently realized a BBR shift uncertainty of $5\times10^{-18}$ \cite{Jefferts2014}, while a Sr optical lattice clock that shuttled the lattice-confined atoms into a cryogenic environment for spectroscopic interrogation achieved an uncertainty of $1\times10^{-18}$ \cite{Ushijima2014}.

In this Letter, we describe the implementation of a thermal radiation `shield' in an Yb optical lattice clock.  This shield enables precise characterization of the BBR environment bathing the ultracold atoms to achieve $1\times10^{-18}$ BBR clock shift uncertainty at room-temperature. By subsequently heating the shield, we directly observe the temperature dependence of the blackbody Stark shift, which corroborates the room-temperature BBR characterization presented here.  The shield also acts as a thermal low-pass filter, protecting the atoms from short-term fluctuations in the blackbody environment that, as lattice clocks continue to improve, could otherwise compromise their stability.  Furthermore, the shield serves as a Faraday enclosure, protecting the atoms from static Stark shifts from stray charges that might accumulatee on the vacuum apparatus \cite{Lodewyck12a}.

\begin{figure}[t]
\includegraphics[width=0.9\linewidth]{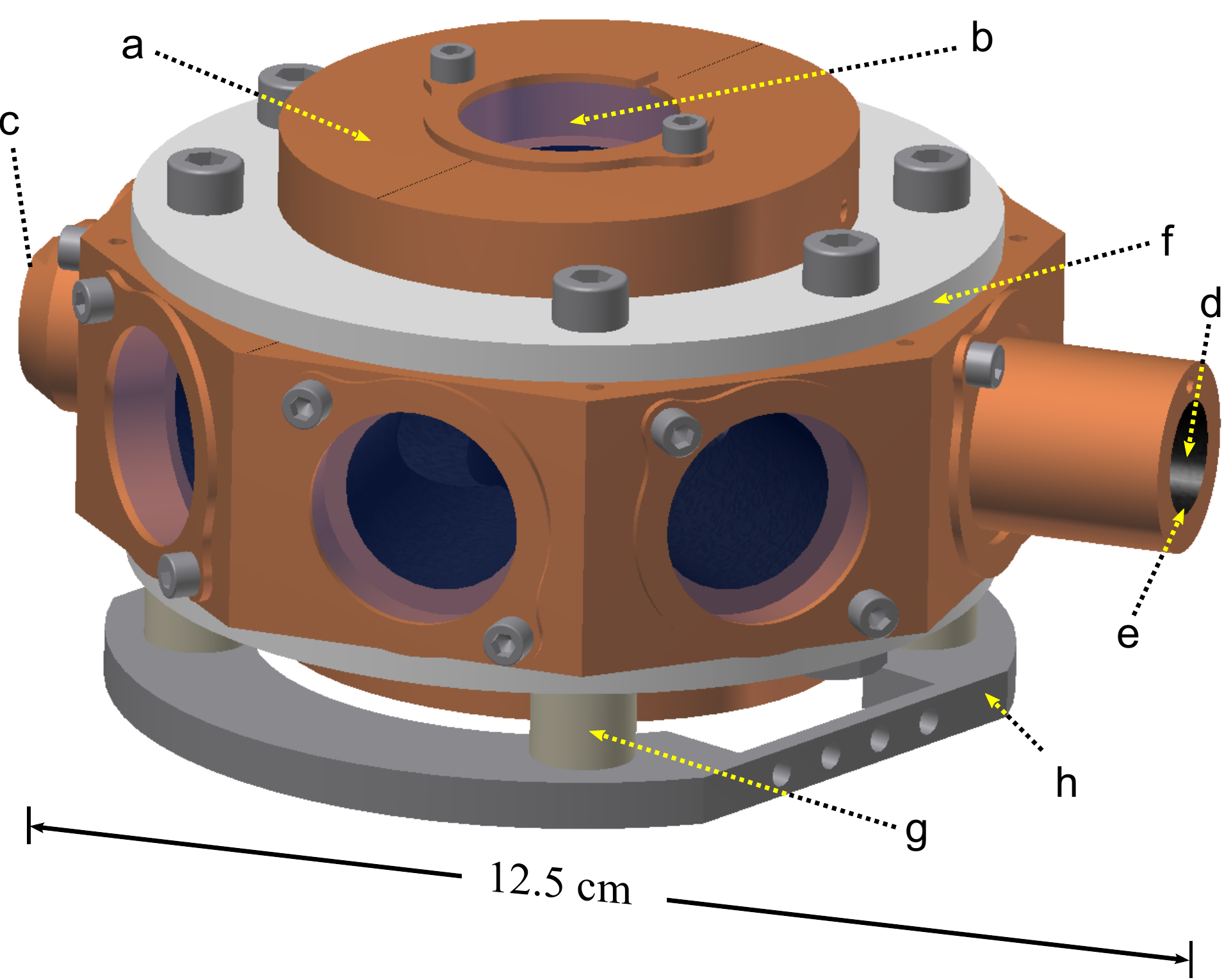}
\caption{(color online) CAD rendering of our in-vacuum radiation shield for an optical lattice clock. Notable features include (a) copper construction, (b) BK7 windows with transparent conductive coatings ($\times7$), (c) atomic beam entry aperture, (d) atomic beam exit aperture, (e) carbon nanotube coating (all internal copper surfaces), (f) boron nitride holding rings ($\times2$), (g) PEEK plastic support posts ($\times4$), and (h) stainless steel support plate. Only the support plate physically contacts the surrounding vacuum chamber.}
\label{Fig:CADshield}
\end{figure}

The BBR shield, shown in Fig.~\ref{Fig:CADshield}, possesses a number of important features to realize these objectives.  The shield demonstrates exceptional temperature uniformity, provided by highly thermally-conductive materials that sit isolated inside ultra-high vacuum (UHV) at $\sim\!2\times10^{-9}$ torr.  Several calibrated platinum resistance temperature detectors (RTDs) distributed throughout the shield provide an accurate, real-time measure of the shield's absolute temperature.  The shield has an important difference from other radiation enclosures, such as those used in blackbody reference sources or to thermally isolate optical interferometers.  It must allow sufficient physical access to perform an atomic physics experiment, requiring careful quantum control and precision measurement, within the well-controlled radiation environment of the enclosure.  Seven windows allow optical access while being nearly opaque to room-temperature BBR. Two apertures on opposing sides of the shield allow a collimated beam of slowed atoms to pass through the central region, providing a source for the lattice-trapped sample. In all, the shield accommodates the collection, cooling and trapping, interrogation, and state-detection of the atoms.  Though the apertures expose the internal volume to BBR from outside the shield, influence of this radiation is minimized with a high-emissivity, carbon nanotube coating applied to all internal surfaces of the shield body.  The shield design and generalized radiation analysis presented below provides a useful framework for other experiments that may require carefully-controlled radiation environments, such as quantum information protocols using Rydberg systems \cite{Saffman2010}.

In normal operation, the shield is passively coupled (albeit weakly) to the surrounding vacuum chamber through conductive and radiative heat transfer, with the combined system near thermal equilibrium at room-temperature. Nevertheless, temperature inhomogeneities exist on some level (e.g., due to local heat sources on the apparatus or drifts in the ambient laboratory temperature).
To account for departures from an isothermal environment, we employ a radiation model capable of capturing the essential physical details. Namely, we model the internal surfaces of the shield and windows as opaque, diffuse, graybody surfaces having temperature-independent emissivities. Motivated by the fact that each aperture opens up to the larger volume of the closed vacuum chamber, we further take the two apertures, as viewed from the inside, to be disk-shaped blackbody surfaces matching the aperture sizes. Collectively these surfaces fully enclose the atoms.
The effective radiation temperature at the center of the shield where the atoms reside, $T_\mathrm{eff}$, is given by the local field energy density, $u$, and can be determined from the radiating surfaces surrounding the atoms:
\begin{equation}
T_\mathrm{eff}^4=
\frac{c}{4\sigma}u=
\sum_i\left(\frac{\Omega^\mathrm{eff}_i}{4\pi}\right)T_i^4,
\label{Eq:Teff}
\end{equation}
where $c$ is the speed of light and $\sigma$ is the Stefan-Boltzmann constant.  The index $i$ runs over all enclosure surfaces, with $T_i$ the temperature and $\Omega^\mathrm{eff}_i$ the effective solid angle of surface $i$. Effective solid angles are non-negative, depend on the geometry and emissivity of all enclosure surfaces, and satisfy the normalization $\sum_i\Omega^\mathrm{eff}_i=4\pi$. In the limit of a completely black (unit-emissivity) enclosure, $\Omega^\mathrm{eff}_i$ reduces to the geometric solid angle subtended by surface $i$ as perceived by the atoms.
More details regarding effective solid angles in the context of BBR clock shifts, including analytical examples, will be presented in a future publication.

\begin{figure}
\includegraphics[width=0.95\linewidth]{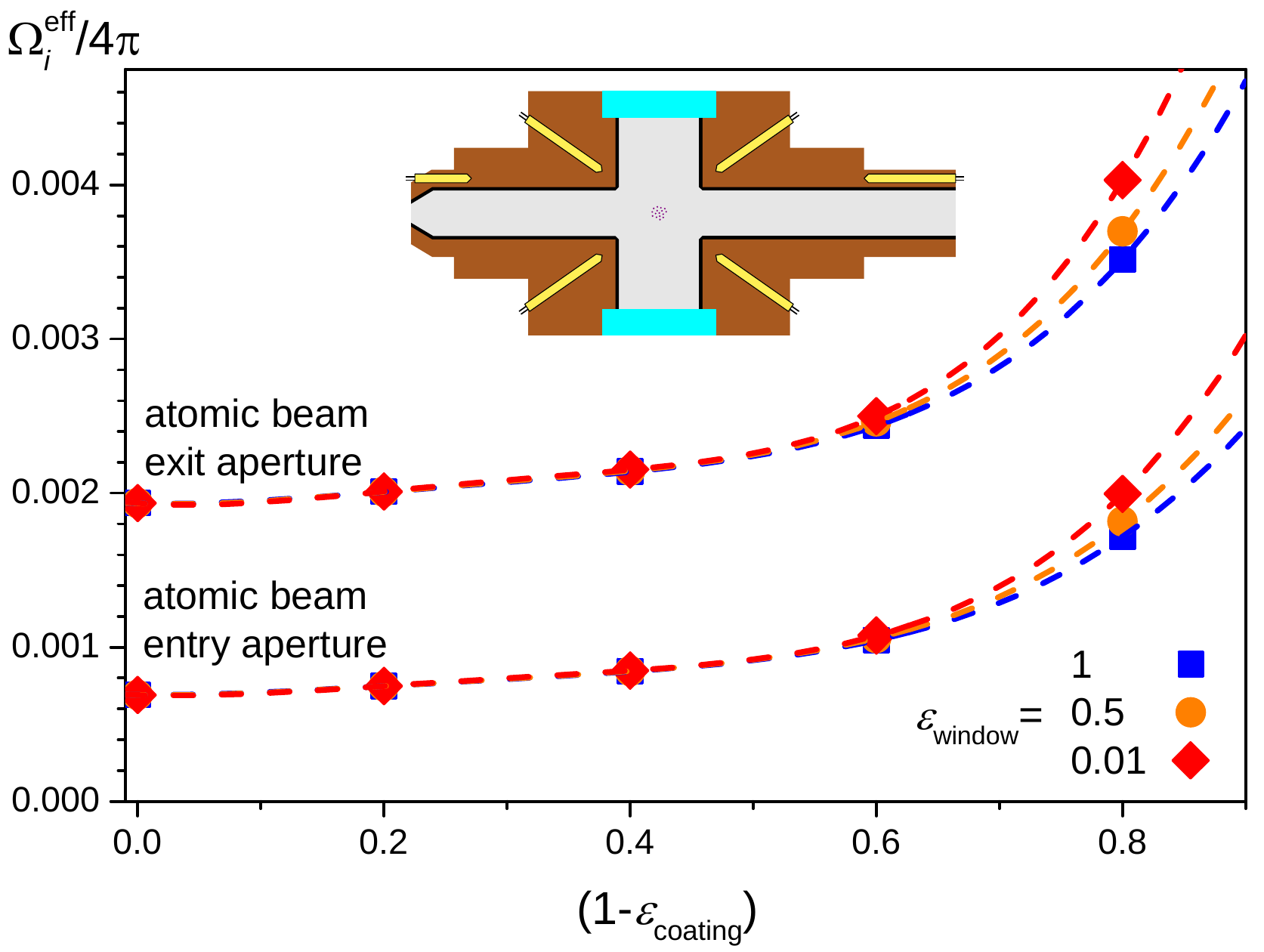}
\caption{(color online) Effective solid angles of the apertures derived from a finite element radiation analysis. $\varepsilon$ denotes emissivity. The inset depicts a two-dimensional cross section of the BBR shield, with the entry and exit apertures to the left and right, respectively.  Top and bottom windows are shown as blue substrates, and yellow cylinders depict the RTDs embedded in the shield.}
\label{Fig:effsa}
\end{figure}

We deduce effective solid angles for our shield enclosure with a finite element (FE) radiation analysis. We supplement the enclosure geometry with a small blackbody sphere, or `probe,' replacing the atoms. The probe's temperature is governed by radiative exchange with the enclosure as modeled by our FE program \cite{ANSYS13}. Associating the probe temperature with $T_\mathrm{eff}$ for different input configurations of the surface temperatures allows extraction of the individual $\Omega^\mathrm{eff}_i$ from Eq.~(\ref{Eq:Teff}). Figure~\ref{Fig:effsa} displays results of our FE analysis, highlighting the apertures specifically. The effective solid angle of each aperture is shown for various combinations of coating and window emissivities. For a perfectly black coating, both the $\Omega^\mathrm{eff}_i$ are independent of the window emissivity and reduce to their respective geometric solid angles. This is the consequence of our design, wherein the apertures are not permitted direct line-of-sight to the windows. As the coating emissivity departs from unity, the $\Omega^\mathrm{eff}_i$ increase while simultaneously acquiring a dependence on window emissivity. However, as Fig.~\ref{Fig:effsa} illustrates, the $\Omega^\mathrm{eff}_i$ are found to be largely constant over a range of moderately-high coating emissivity. Here we find a two-fold advantage in using a coating of high-emissivity: (i) it minimizes the overall influence of BBR entering through the apertures and (ii) it minimizes the corresponding sensitivity to the precise emissivity values.  The high emissivity coating that we employ consists of multi-wall carbon nanotubes \cite{Lehman2011} applied to the interior shield body surfaces \cite{Lehman2007}. The coating is highly-thermally and -electrically conductive and, most importantly, exhibits a high surface emissivity. Hemispherical reflectance measurements indicate emissivity $\varepsilon_{\mathrm{coating}}>0.8$, measured for wavelengths from visible to 20 $\mu$m. Taken together with measurements of $\varepsilon_{\mathrm{window}}$, in practice we can equate effective solid angles to geometric solid angles, with the FE analysis providing a means to gauge corresponding uncertainties.

Special effort must be made to mitigate potential error in the shield temperature measurement. After a thermal-cycling process, a NIST-traceable absolute temperature calibration is performed on each RTD by the manufacturer. The simultaneous use of many (seven) sensors for temperature measurement of the shield aids in the detection of calibration shifts that can sometimes occur.  Self-heating from ohmic dissipation of the RTD sense current can be meaningful, especially in vacuum. We have directly measured self-heating of $\sim 7$ K/mW, which for our low sense current (96 $\mu$A into 109 $\Omega$) yields an effect of 7 mK.  Vacuum also reduces thermal transfer from the platinum wires to the RTD ceramic housing, enabling parasitic heat flow through the RTD leads.  To counter this effect, a thermally-conductive (but electrically-insulative) epoxy covers the entire RTD and its leads, making excellent thermal contact between the shield and all parts of the sensor.

Unlike the copper shield, the temperature of the windows is not directly measured in real-time.  Because of finite thermal contact between the window substrates and copper shield (mediated by a thin carbon-loaded polyimide layer), the windows may exhibit small temperature deviations from the rest of the shield.  This effect was assessed by temporarily fitting a window with a temperature sensor and heater.  The measured temperature difference between the window and shield as a function of heater power indicated the thermal conductance from the window.  In the clock apparatus, this leads to a clock uncertainty of $3 \times 10^{-19}$.  The BK7 windows are also mostly transparent to radiation below 3 $\mu$m and weakly transparent ($<\!1\%$) above 3 $\mu$m.  This enables a small fraction of BBR to enter or escape through the windows.  Here we benefit from room temperature operation: the radiative correction associated with the windows' partial transparency is negligible because both the shield and the surrounding apparatus are approximately the same temperature.

Table \ref{Tab:errorbudget} summarizes the BBR shift uncertainties.  In addition to items that have been described so far, we note that variations in the position of the lattice-trapped atoms from the geometric center of the chamber, non-scalar Stark shifts from anisotropy in the BBR, and the application of $T_\mathrm{eff}$ in the dynamic correction all lead to comparatively small uncertainties.  The total uncertainty associated with the BBR environment is $5.5 \times 10^{-19}$.  The last four items in the upper portion of Table \ref{Tab:errorbudget} list uncertainties stemming from atomic response.  The mean and uncertainty of the dynamic correction $\eta_1$ is taken as the weighted average of three distinct determinations of its value \cite{Beloy2012,Safronova2012}. Magnetic dipole ($M1$) interaction with the BBR leads to a $\sim\!3\times10^{-20}$ clock shift, with a small uncertainty included for the response factor. Combining the uncertainty from the BBR environment and atomic response yields a total uncertainty for the BBR shift, Eq.~(\ref{Eq:DeltanuBBR}), of $1 \times 10^{-18}$.

\begin{table}
\caption{BBR shift uncertainty ($\times10^{-19}$ clock frequency) for our Yb lattice clock in normal operation ($\sim\!296.7$ K).}
\label{Tab:errorbudget}
\begin{ruledtabular}
\begin{tabular}{lc}
\multicolumn{2}{c}{{\it BBR environment}}\\
RTD temperature measurements						\\
\quad manufacturer calibration (5 mK)				& $1.6$ \\
\quad post-calibration fidelity						& $2.4$ \\
\quad digital multimeter (4-wire)					& $2.2$ \\
\quad self-heating									& $1.6$ \\
\quad parasitic conduction/radiation				& $0$ \\
temperature inhomogeneity/effective solid angles \\
\quad BK7 windows									& $2.9$ \\
\quad entry aperture (oven shielded by shutter)		& $2.4$ \\
\quad exit aperture									& $0.3$ \\
other \\
\quad application of $T_\mathrm{eff}$ in dynamic correction	& $0.1$ \\
\quad residual transmission through windows			& $0.2$ \\
\quad atomic position/dimensional tolerances		& $0.5$ \\
\quad BBR anisotropy (non-scalar Stark)				& $0$ \\
\multicolumn{2}{c}{{\it atomic response}}\\
differential static polarizability					& $0.5$ \\
dynamic correction $\eta_1$      					& $8.5$ \\
dynamic correction $\eta_2$                         & $0.4$ \\
BBR Zeeman ($M1$) factor							& $0.1$ \\
\hline
Total BBR environment				& $5.5$ \\
Total atomic response				& $8.5$ \\
Total 								& $10$
\end{tabular}
\end{ruledtabular}
\end{table}

\begin{figure}
\includegraphics[width=1.0\linewidth]{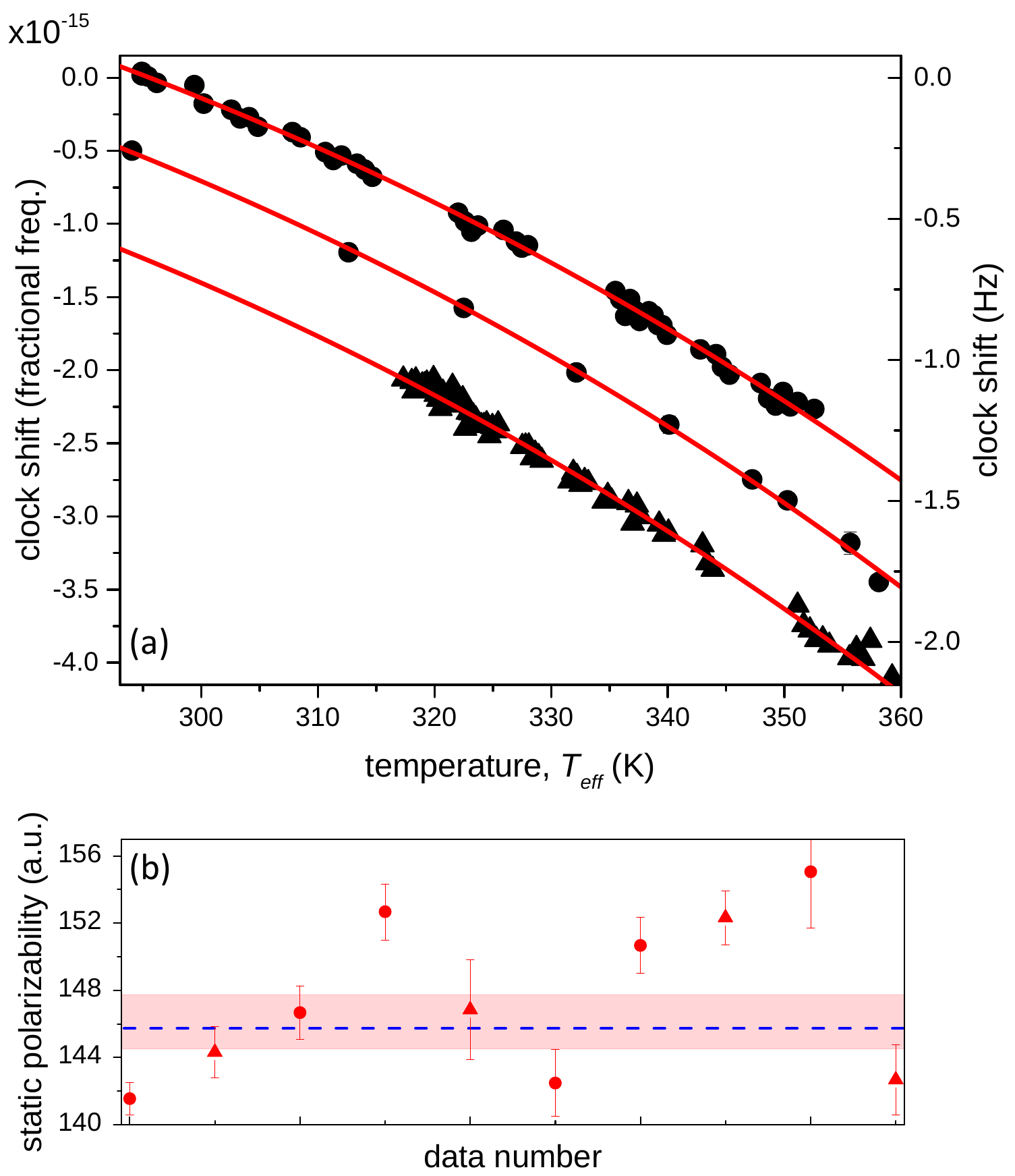}
\caption{(color online) Measured BBR Stark shift versus temperature.  (a) The case of heating (dots) and cooling (triangles), each with a fit (red solid curve) using Eq.~(\ref{Eq:DeltanuBBR}).  The curves are intentionally offset from one another for visual clarity. (b) The result of ten measurements of the BBR shift versus temperature.  The differential static polarizability, $\Delta \alpha(0)$, is extracted from each fit and shown here.  Circles denote measurements while heating the shield, whereas triangles denote measurements while letting the shield cool.  The shaded region denotes $\pm\!\,1\sigma$ weighted standard error \cite{BevingtonBook} of the measured $\Delta \alpha(0)$ values.  The dashed blue line gives the expected result \cite{Sherman2012}.}
\label{Fig:shiftvstempdata}
\end{figure}

Since the determination of the blackbody environment plays such a critical role in the final uncertainty budget of an optical lattice clock, it is imperative to experimentally validate that determination, in this case characterized by $T_\mathrm{eff}$. To this end, we heat the BBR shield to directly observe the temperature dependence of the BBR shift.  On the one hand, this measurement could be used to determine the atomic response parameters, $\Delta \alpha(0)$ and $\eta_{1,2}$ in Eq.~(\ref{Eq:DeltanuBBR}). However, since these parameters have been independently determined to a high level of accuracy, here it is more meaningful to compare the measured and expected BBR shift temperature dependence as a consistency check on our determination of $T_\mathrm{eff}$.  We fit annulus-shaped resistive heaters on the top and bottom of the shield (nested below the boron-nitride holding rings).  This enables us to raise the shield temperature by up to $100$ K above room temperature during operation of the lattice clock, limited only by vacuum degradation of the trap lifetime at the highest temperatures.  We operate two Yb lattice clocks and make direct frequency measurements between them.  One lattice clock is fitted with the BBR shield and heaters, while the second serves as an optical frequency reference. The uncharacterized ambient BBR environment within the second system is known to be sufficiently stable over the course of a measurement (several hours) \cite{Hinkley2013}.

While comparing the atomic clock frequencies, we gradually raise the temperature of the BBR shield of the first clock and then allow the shield to cool to room temperature ($1/e$ time of $\sim\!3$ hours).  The observed clock shift versus temperature is plotted in Figure \ref{Fig:shiftvstempdata}(a).  The results from three distinct measurement protocols are shown: the top curve shows measurement for the case of a slow continuous heating of the shield temperature, the middle curve for the case of controlled intermittent heating to allow the shield to settle at a nearly-constant temperature for each measurement point, and the bottom curve for the case of passive cooling of the shield after a heating cycle.  Whatever measurement protocol was employed, we ensured that temperature changes were sufficiently slow to avoid any meaningful Doppler shifts from optical phase chirps caused by thermal expansion of the windows or temperature dependence of the index of refraction.  While the shield body is heated and subsequently cooled, the apertures expose the atoms to unchanging room temperature BBR.  The temperature of the shield windows closely follows that of the shield body, with a difference determined by thermal conductance measurements described above together with the estimated radiative heat transfer from its surfaces.  Red solid curves fit the data to Eq.~(\ref{Eq:DeltanuBBR}).  Here, the differential static polarizability, $\Delta \alpha(0)$, is the fit parameter and the known dynamic corrections for the Yb lattice clock, $\eta_1=0.01745(38)$ and $\eta_2=0.000593(16)$, are fixed.  Figure \ref{Fig:shiftvstempdata}(b) shows the results of ten distinct measurements of the BBR shift temperature dependence.  The weighted mean of the measured differential static polarizability is found to be $\Delta \alpha(0) = 146.1(1.3)$ a.u., in excellent agreement with static Stark measurements, $\Delta \alpha(0) = 145.726(3)~\mathrm{a.u.}=h \times 3.62612(7) \times 10^{-6}$ Hz/(V/m)$^2$ ~\cite{Sherman2012}.  This agreement with an independent, fully-constrained measurement of the atomic response parameters provides validation of our determination of $T_\mathrm{eff}$, the first such validation for an optical lattice clock.

In conclusion, we have demonstrated a room-temperature radiation shield in an ytterbium optical lattice clock.  The resulting BBR shift uncertainty from the thermal environment is $5.5 \times 10^{-19}$.  The total uncertainty for the BBR shift, the largest uncanceled shift in the optical lattice clock, is $1 \times 10^{-18}$. We note that this level of control is achieved with the simplicity of room-temperature operation and without requiring the special transport of lattice trapped atoms to a cryogenic environment. Moreover, our shield design is expected to be immediately applicable to optical lattice clocks based on other atomic species. Replacing Yb inside our shield with Mg, Ca, Sr, or Hg, for example, the uncertainty from the BBR environment would be $9\times10^{-20}$, $6\times10^{-19}$, $1\times10^{-18}$, or $4\times10^{-20}$, respectively \cite{Porsev2006,Hachisu2008}. Because the BBR environment uncertainty has now been significantly reduced, despite recent measurements and calculations, the dynamic correction now dominates the BBR shift uncertainty, inviting further investigation. Our explicit measurement of the BBR shift temperature dependence supports the analysis presented here.  This work represents a key step towards realizing an optical lattice clock with total uncertainty of $1 \times 10^{-18}$, enabling a variety of fundamental physics measurements at unprecedented levels of precision \cite{Schiller2009,Rosenband2008,Derevianko2014,Arvanitaki2014}.

This work was supported by NIST, DARPA QuASAR, NASA Fundamental Physics, DARPA PULSE, and NRC-RAP.  We thank W.~Tew, G.~Strouse, and D.~Cross of NIST for useful discussions on precision temperature measurement and W.~McGrew and S.~Jefferts for careful reading of the manuscript.


\end{document}